\documentstyle[epsfig,prl,twocolumn,aps]{revtex}
\begin{document}
%\draft
\newcommand{\vek}[1]{{\mbox{\boldmath $#1$}}}
%\newcommand{\comment}[1]{{\mbox{\ \ \ $\vdots$[{\it \  #1 }]$\vdots$ \ \ }}}
%\tighten
%\preprint{\vbox{\hbox{\large FinaldraftPRL}
%}}

\title{Edge Electron Gas}

\author{Walter Kohn and Ann E. Mattsson}

\address{Department of Physics, UCSB, CA 91306-9530}

%\date{\today}

\author{
%\begin{abstract}
\widetext\leftskip=0.10753\textwidth \rightskip\leftskip
The uniform electron gas, the traditional starting point for density-based many-body theories
of inhomogeneous systems, is inappropriate near electronic edges. In its place we put forward the 
appropriate concept of the edge electron gas.\hfill
%\end{abstract}
}

\maketitle

%\pacs{71.15.Mb,31.15.Bs,31.15.Ew,73.90+f}

%\newpage
\narrowtext
%%%%%%%%%%%%%%%%%%%%%%%%%
%%%%%%%%%%%%%%%%%%%%%%%%%%
Since the work of Thomas and Fermi (TF)~\cite{TF} through the 
earliest papers on density functional theory (DFT)~\cite{DFTorg} up 
to ongoing work involving density gradients~\cite{GGA}, the {\em uniform electron gas} 
has been the starting 
point of density-based approximate theories of inhomogeneous systems. However, 
real physical systems have {\em electronic 
edge regions} where effective single particle wave functions 
evolve from oscillatory to evanescent, and where clearly approximations starting from a uniform 
electron gas are ill-founded.   

\begin{figure}
\begin{center}
\mbox{\epsfig{file=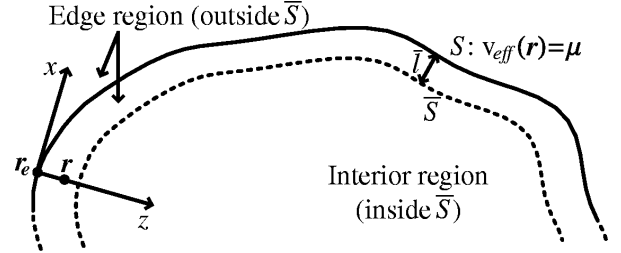,width=3.1in,angle=0}}
\end{center}
\caption{Interior and Edge regions of a bounded system. $S$ is the nominal dividing surface (-----).
$\bar{S}$ (- - -) is the physical dividing surface with $\bar{l}=\gamma l$. 
$l$ is defind in Eq.~(\protect\ref{eq:ledef}) and $1<\gamma<3$.} \label{fig:edgeblob}
\end{figure}

The {\em edge surface} $S$, for a given system, may be precisely 
defined by 
%%%%%%%%%%
\begin{equation}
v_{eff} (\vek{r}) = \mu \, , 
\end{equation}
%%%%%%%%%%%%%%%
where $v_{eff} (\vek{r})$ is the exact self-consistent 
potential of the Kohn-Sham (KS) equations~\cite{DFTorg} and $\mu$ is the chemical potential, 
half-way between the highest 
occupied and the lowest unoccupied KS orbital energies (In what follows we take $\mu=0$). 
Outside of $S$ all occupied  KS orbitals {\em decay exponentially} (See Fig.~\ref{fig:edgeblob}).

Much interesting science involves 
precisely such edge regions (See Fig.~\ref{fig:edgeblob}), 
such as ionization energies of the outermost electrons,
molecular binding, surface energies etc. For these regions, in place of the uniform electron gas,
we discuss in this paper the new concept 
of the {\em edge electron gas}, which {\em is} valid 
in the edge region near $S$~\cite{KohnAG}.

The edge gas concept is based on the principle of ``nearsightedness'' put forward in~\cite{near}:
The local electronic structure near a point $\vek{r}$, while strictly speaking requiring a knowledge 
of the density $n(\vek{r}')$, or equivalently of the effective potential $v_{eff} (\vek{r}')$, 
{\em everywhere}, in fact is largely determined by $v_{eff} (\vek{r}')$ for $\vek{r}'$ {\em near} 
$\vek{r}$. Nearsightedness applies just as much to the edge region as to the interior region far 
inside $S$.

This principle has been applied to the density $n(\vek{r})$ and to the one particle density 
matrix~\cite{near}. In the present paper we also apply it to the 
exchange energy $E_{x}$ as follows: We write
%%%%%%%eq%%%%%%%%
\begin{equation}
E_{x}=- \frac{e^2}{2} \int n(\vek{r}) R^{-1}_{x}(\vek{r})\, d\vek{r} \, .
\end{equation}
%%%%%%%endeq%%%%%%%
where $R_{x}^{-1}(\vek{r})$, the inverse radius of the exchange hole, is defined by
%%%%%%%eq%%%%%%%%
\begin{equation}
R_{x}^{-1}(\vek{r}) \equiv -\int \frac{n_{x}(\vek{r};\vek{r}')}{|\vek{r}-\vek{r}'|} d\vek{r}' \, ,
\label{eq:3Dinvraddef}
\end{equation}
%%%%%%%endeq%%%%%%%
and $n_{x}(\vek{r};\vek{r}')$ is the exchange hole density which satisfies
$\int n_{x}(\vek{r};\vek{r}') d\vek{r}' = -1 $.

%%%%%%fig%%%%%%%%%%
\begin{figure}
\begin{center}
\mbox{\epsfig{file=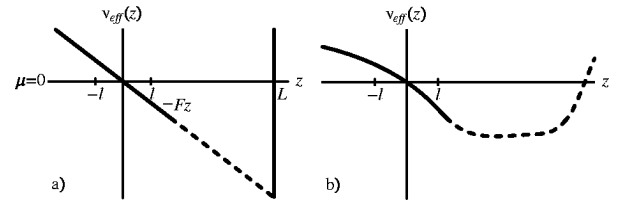,height=3.1in,angle=270}}
\end{center}
\caption{a) The Airy gas: a strictly
linear $v_{eff}$, with a hard wall at the distant point $L$. b) The edge gas: a typical 
$v_{eff}(z)$ near the edge at $z=0$. All levels up to $\mu=0$ are occupied.} \label{fig:edgesystems}
\end{figure}

%%%%%endfig%%%%%%%%%%%
For a consideration of the electronic structure at a point $\vek{r}$ near $S$ we first drop a 
perpendicular to the nearest point $\vek{r}_e$ on $S$, take $\vek{r}_e$ as origin of 
coordinates and the 
$z$-axis through $\vek{r}$ (See Fig.~\ref{fig:edgeblob}.); thus $\vek{r}=(0,0,z)$.
Near $\vek{r}_e$ we have, up to first order in $(x,y,z)$,
%%%%%%%%eq%%%%%%%%%%%%%%
\begin{equation}
v_{eff} (\vek{r})= -F z \, . \ \ \ \ \ \ \mbox{(Airy Gas)}
\end{equation}
%%%%%%%%%%%endeq%%%%%%%%%%%%%%
This leads to the concept of the  Airy gas (AG), the simplest version of the edge gas 
(Fig.~\ref{fig:edgesystems}a). The AG is a 
principal subject of this paper. Of course only the properties of the AG near the edge at 
$z=0$ are of physical interest.
%%%%%%fig%%%%%%%%%%
%%%%%endfig%%%%%%%%%%%
In many cases $v_{eff}$ varies rapidly along $z$ but slowly 
along $x$ and $y$, i.\ e.\ $\kappa l \mbox{\scriptsize \  $\stackrel{ <}{ \sim}$ \ } 1$, where 
$l$ is the nominal surface ``thickness'', defined in Eq.~(\ref{eq:ledef}) below, and 
$\kappa$ is the curvature of $S$ at $\vek{r}_e$. This leads to the 
more general concept of the 
Edge Gas (EG),
%%%%%%%%eq%%%%%%%%%%%%%%
\begin{equation}
v_{eff} (\vek{r})= v_{eff}(z)\, ;\ \  v'_{eff}(0)=-F \, , \ \  \mbox{(Edge Gas)}
\end{equation}
%%%%%%%%%%%endeq%%%%%%%%%%%%%%
(Fig.~\ref{fig:edgesystems}b).
The EG can be refined by including curvature corrections~\cite{KohnPicus}.

There have been extensive previous studies of several model systems, 
notably by V.\ Sahni, A.\ G.\ Eguiliz and their collaborators~\cite{SahniEguiliz++}, of 
electrons moving in effective potentials $v_{eff}(z)$ with an electronic edge, 
say at $z=0$. Their common features are: $v_{eff}(-\infty)>\mu$ and 
$v_{eff}(+\infty)=\mbox{constant}<\mu$. 
The differences between earlier work and the present contribution are: (i) Our emphasis on the 
{\em locality} of the electronic edge structure, i.\ e.\ its independence of the behavior of 
$v_{eff}(\vek{r})$ far from the edge; (ii) our detailed analysis of the simplest universal 
edge model, 
$v_{eff}(z)=-z \ \ \  (-\infty<z<+\infty)$ which, by appropriate scaling, provides a 
first approximation 
for dealing with general $3D$ systems with edges; (iii) preliminary discussions of further 
improvments.

%%%%%%%%%%%%%%%%%%%%%%%%%%%%%%%%%%%%%%%%%%%
The Edge Electron Gas:
We now consider the situation in Fig.~\ref{fig:edgesystems}b where the effective potential is 
independet of $x$ and $y$ but generic in $z$.

The surface ``thickness'' $l$ and the corresponding energy $\epsilon$ are defined uniquely, 
up to constant factors, by
%%%%%%%%%%%%%%
\begin{equation}
l \equiv \left(\frac{\hbar^2}{2 m F}\right)^{1/3}\, ; \ \ \  
\epsilon \equiv \left(\frac{\hbar^2 F^2}{2 m}\right)^{1/3}\, ,
\label{eq:ledef}
\end{equation}
%%%%%%%%%%%%%%%%
where $F \equiv | v'_{eff}(0) | $.

The normalized eigenfunctions of the KS equations~\cite{DFTorg} are
%%%%%eq%%%%%%%%
\begin{equation}
\psi_\nu (x,y,z) = \varphi_j (z) \frac{1}{A^{1/2}} e^{i(k_1 x + k_2 y)}\, ; \ 
(\varphi_j,\varphi_{j'})=\delta_{j \, j'}
\label{eq:edgewavefunc}
\end{equation}
%%%endeq%%%%%%%
where $\nu \equiv (j,k_1,k_2)$; $k_i L_i = 2 \pi m_i \ (i=1,2)$ and $A \equiv L_1 L_2$ 
is the cross-sectional area.
The functions  $\varphi_j (z)$ satisfy
%%%eq%%%%%%%
\begin{equation}
(-\frac{\hbar^2}{2 m} \frac{d^2}{dz^2} + v_{eff}(z) - \epsilon_j) \varphi_j (z) = 0
\end{equation}
%%endeq%%%%%%%%%
and the boundary conditions
%%%%eq%%%%%%%%%%
\begin{equation}
\varphi_j (-\infty)=\varphi_j (+ \infty)=0 \ \ ,
\end{equation}
%%%%%%%%endeq%%%%%%%
The $\epsilon_j$ are negative and, for each $\epsilon_j$,  
$(k_1,k_2)$ satisfy the inequality
%%%%%%%%%%eq%%%%%%%%%%%
\begin{equation}
\epsilon_\nu \equiv \frac{\hbar^2}{2 m} (k_1^2+k_2^2) +\epsilon_j \leq 0 \ \ .
\end{equation}
%%%%%%%%%endeq%%%%%%%%%%%

The density of the EG is given by
%%%%%%%%%%eq%%%%%%%%%%
\begin{equation}
n(z)=2 \sum_j \varphi_j^2 (z) w_j \, ; \ \ \  w_j=\frac{m}{2 \pi \hbar^2} |\epsilon_j |
\label{eq:EGdensity}
\end{equation}
%%%%%%%%%%%%endeq%%%%%%%%%%%
where the factor $2$ accounts for the spin and where $w_j$
is the weight factor accounting for the number of occupied wave-vectors $(k_1,k_2)$.

The exchange hole density, $n_x(\vek{r};\vek{r}')$, is defined by 
%%%%%%%eq%%%%%%%%
\begin{eqnarray}
n_x(\vek{r};\vek{r}')  &\equiv&  
- \frac{1}{2} \frac{| \rho_1(\vek{r};\vek{r}')|^2}{n(\vek{r})}\, ; \label{eq:holedef} \\
\rho_1(\vek{r};\vek{r}')   &\equiv&  2 \sum_{\epsilon_\nu \leq 0} \psi_\nu(\vek{r}) 
\psi_\nu^*(\vek{r}') \, . \label{eq:rho1def}
\end{eqnarray}
%%%%%%%endeq%%%%%%%

\begin{figure}
\begin{center}
\mbox{\epsfig{file=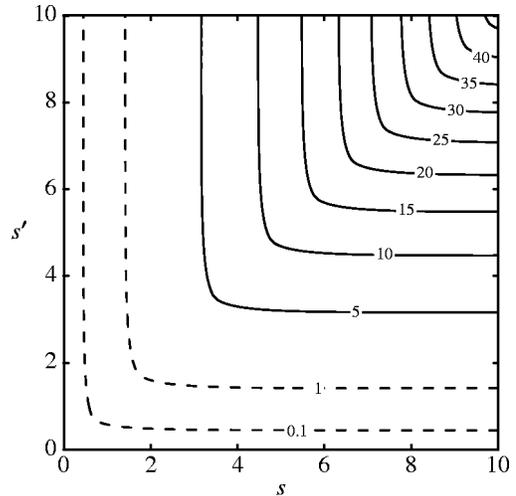,height=2.6in,angle=270}}
\end{center}
\caption{The universal function  $g(s,s')$ of Eq.~(\protect\ref{eq:gfunc}).}
\label{fig:gfunc}
\end{figure}

Inserting Eq.~(\ref{eq:edgewavefunc}) in Eq.~(\ref{eq:rho1def}) and integrating over $k_1$ and 
$k_2$ gives
%%%%%%%eq%%%%%%%%
\begin{equation}
\rho_1(\vek{r};\vek{r}')= \frac{1}{\pi} \sum_j \varphi_j (z) \varphi_j (z') 
\frac{k_j J_1(k_j \rho')}{\rho'}  \ \ 
\label{eq:rho1EG}
\end{equation}
%%%%%%%endeq%%%%%%%
where $J_1(x)$ is the first order Bessel function and where, without loss of generality, 
we have set $\vek{r}=(0,0,z)$, and have expressed $\vek{r}'$
in cylindrical coordinates $\rho', \alpha'$ and $ z'$. 
$k_j$ is the maximum transverse wave number associated with $\varphi_j (z)$ given by
$k_j=({2 m | \epsilon_j |}/{\hbar^2})^{1/2}$.
This leads to 
%%%%%%%eq%%%%%%%%
\begin{eqnarray}
R_x^{-1}(\vek{r})&=&\frac{1}{\pi\  n(z)} \int dz' \sum_j \sum_{j'} \varphi_j (z) \varphi_j (z') 
\varphi_{j'} (z) \varphi_{j'} (z') \times \nonumber \\
&& \hspace{0.8in} \times {(\Delta z)}^{-3} \, g(k_j \Delta z , k_{j'}\Delta z) \ \ , 
\label{eq:invraddef}
\end{eqnarray}
%%%%%%%endeq%%%%%%%
where $\Delta z = |z-z'|$ and the universal function $g(s,s')$, independent of $v_{eff}(z)$, is 
given by
%%%%%%%eq%%%%%%%%
\begin{equation}
g(s,s') \equiv 
s \, s' \int_0^\infty dt 
\frac{J_1(s t) J_1(s' t)}{t \sqrt{1 + {t}^2}} \ \ .
\label{eq:gfunc}
\end{equation}
%%%%%%%endeq%%%%%%%
We have calculated $g(s,s')$ numerically and present a contour plot in Fig.~\ref{fig:gfunc}.
Further details about $g(s,s')$ are available from A.\ E.\ Mattsson.
%%%%%%%%%%%fig%%%%%%%%%%%
%%%%%%%%%%endfig%%%%%%%%%%%%%

\begin{figure}
\begin{center}
\mbox{\epsfig{file=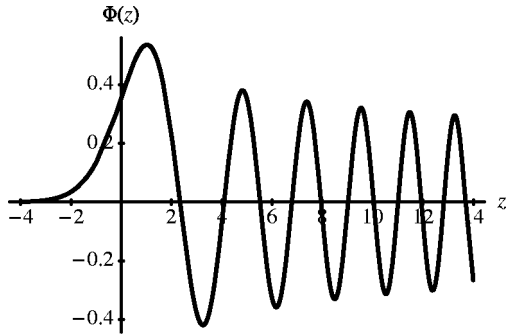,height=2.6in,angle=270}}
\end{center}
\caption{The function $\Phi(z) \equiv Ai(-z)$\protect\cite{HMF}.}
\label{fig:Airy}
\end{figure}

The Airy Gas:
We now consider the simplest model of an edge electron gas, the {\em Airy Gas} (AG)
of Fig.~\ref{fig:edgesystems}a, in which the potential has the strictly 
linear form 
%%%%%%%%%eq%%%%%%%
\begin{equation}
v_{eff}(z)= \left\{ \begin{array}{lcl}
             -F z  & -\infty<z<L & (F >0)\\
             +\infty & z \geq L  & (L/l \rightarrow \infty)
             \end{array} \right. \ \ .
\end{equation}
%%%%%%%%endeq%%%%%%%%%
The normalized eigenfunctions $\varphi_j (z)$ satisfy
%%%%%%eq%%%%
\begin{eqnarray}
&&(-\frac{\hbar^2}{2 m} \frac{d^2}{dz^2} -F z - \epsilon_j)\varphi_j (z) =0\, , \\ 
&&\varphi_j (-\infty)=\varphi_j (L)=0 .
\end{eqnarray}
%%endeq%%%%%%%%%
In suitable variables {\em all} the $\varphi_j (z)$ are proportional to the Airy function
$Ai(-z)$; Ai(z) satisfies 
%%%%%%eq%%%%
\begin{equation}
(- \frac{d^2}{dz^2} + z) Ai (z) = 0 \ \ , \ \ \ \  Ai(+\infty)=0 \ \ .
\end{equation}
%%endeq%%%%%%%%%
The function $\Phi(z)\equiv Ai(-z)$ is shown in Fig.~\ref{fig:Airy}.
%%%%%%%%%%%fig%%%%%%%%%%%
%%%%%%%%%%endfig%%%%%%%%%%%%%
Relevant properties of $\Phi(z)$ are~\cite{HMF}:
%%%%%%eq%%%%
\begin{equation}
\left\{ \begin{array}{l}
      \Phi(0)=0.35503; \ \ \ \ \  \Phi'(0)=0.25882 \\
      z \to -\infty: \Phi(z) \approx \frac{1}{2 \pi^{1/2}|z|^{1/4}} e^{-\frac{2}{3} |z|^{3/2}} \\ 
      z \to +\infty: \Phi(z) \approx \frac{1}{\pi^{1/2}z^{1/4}} 
                       \sin{(\frac{2}{3} |z|^{3/2}+\frac{\pi}{4})}
      
\end{array} \right.
\end{equation}
%%%%%%%%endeq%%%%%%%%%%
In the AG all $3D$ states with energy up to $\mu=0$ are occupied 
(Fig.~\ref{fig:edgesystems}a). The AG is completely characterized 
by the length $l$ and the energy $\epsilon$ in Eq.~(\ref{eq:ledef}).
In units of $l$ and $\epsilon$ all AG's are identical. We shall later verify that for $z >> l$
the AG behaves like a locally homogeneous gas whose density is
$n(z)=(3 \pi^2)^{-1} (2 m F/\hbar^2)^{3/2} z^{3/2}$. Thus the ``thickness'' of the edge
region is $ \sim l$. The $\varphi_j(z)$ contributing to the edge electron density 
have $\epsilon_j$ of the order $-\epsilon$, much less, in absolute value, than the lowest eigenvalue 
($\sim -FL$). These 
small $\epsilon_j$, obtainable from the boundary condition $\varphi_j (L)=0$, and the $\varphi_j(z)$  
are given by
%%%%%%%%%%%%%%%%%%
\begin{equation}
\epsilon_j = -j (\frac{l}{L})^{1/2} \pi \epsilon \, ; \ \
\varphi_j (z)= \frac{\pi^{1/2}}{(L l)^{-1/4}} \Phi(\frac{z}{l} + \frac{\epsilon_j}{\epsilon})\, .
\label{eq:epsphidef}
\end{equation}
%%%%%%%%%%%%%%
We note that the $\epsilon_j$ are equally spaced and that 
successive $\varphi_j (z)$'s are identical except for a small shift, 
$ l (l/L)^{1/2} \pi$, to the right.

Using Eqs.~(\ref{eq:epsphidef}) in Eq.~(\ref{eq:EGdensity}) we obtain for the density
%%%%%%%eq%%%%%%%%%%                                                       
\begin{eqnarray}
n(z)&=&l^{-3} n_0(\zeta)\, , \ \ \ (\zeta\equiv z/l) \\  
n_0(\zeta) &\equiv& \frac{1}{2 \pi} \int_0^\infty \Phi^2(\zeta-\zeta')\zeta' d\zeta' \, .
\label{eq:n}
\end{eqnarray}
%%%%%%%%%%%%endeq%%%%%%%%%%%
%%%%%%%%%%%fig%%%%%%%%%%%
\begin{figure}
\begin{center}
\mbox{\epsfig{file=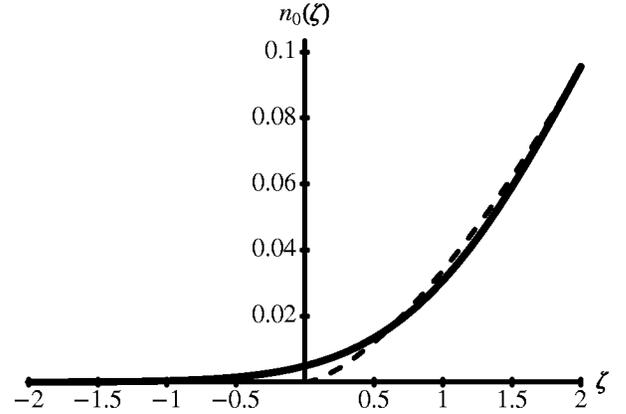,height=3.1in,angle=270}}
\end{center}
\caption{The density of the Airy gas, exact (-----) and in the Thomas Fermi approximation (- - -).}
\label{fig:density}
\end{figure}
%%%%%%%%%%endfig%%%%%%%%%%%%%
Fig.~\ref{fig:density} shows the exponential decay for $\zeta << -1$. 
For $\zeta >> 1$
%%%%%%%eq%%%%%%%%%%                                                       
\begin{equation}
n_0(\zeta) = \frac{1}{3 \pi^2} \zeta^{3/2} - \frac{1}{16 \pi^2} \zeta^{-3/2} 
\sin{(\frac{4}{3} \zeta^{3/2})}+ \cdots \ \ .
\end{equation}
%%%%%%%%%%%%endeq%%%%%%%%%%%
The first term is identical with the Thomas Fermi result (See Fig.~\ref{fig:density}); the second 
describes Friedel oscillations, due to the edge at $z=0$. 

\begin{figure}
\begin{center}
\mbox{\epsfig{file=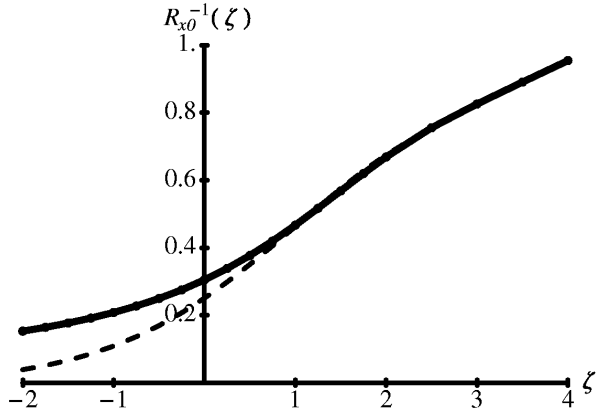,width=3.1in,angle=0}}
\end{center}
\caption{The inverse radius of the exchange hole of the Airy Gas: (-----) exact, (- - -) LDA.}
\label{fig:exchange}
\end{figure}

Exchange:
Inserting the appropriate expressions for the AG, Eqs.~(\ref{eq:epsphidef}), 
into Eq.~(\ref{eq:invraddef}) gives
%%%%%%%eq%%%%%%%%
\begin{equation}
R_x^{-1}(z)= l^{-1} R_{x0}^{-1}(\zeta)  \ \ ,
\end{equation}
%%%%%%%endeq%%%%%%%
where
%%%%%%%eq%%%%%%%%
\begin{eqnarray}
\FL && R_{x0}^{-1}(\zeta)=\frac{1}{\pi\  n_0(\zeta)} \times \nonumber \\
\FL &&\int\limits_{-\infty}^{\infty} {d\zeta'} \int\limits_0^\infty {d\epsilon} 
\int\limits_0^\infty {d\epsilon'} \Phi(\zeta-\epsilon) \Phi(\zeta'-\epsilon)  
\Phi(\zeta-\epsilon') \Phi(\zeta'-\epsilon') \times \nonumber \\
\FL && \hspace{0.85in} \times {|\zeta-\zeta'|}^{-3} 
g(\sqrt{\epsilon}|\zeta-\zeta'|,\sqrt{\epsilon'}|\zeta-\zeta'|) \ \ .
\end{eqnarray}
%%%%%%%endeq%%%%%%%
In Fig.~\ref{fig:exchange} we see the exact $R_{x0}^{-1}(\zeta)$ for the AG compared with its
local density approximation 
(LDA) result 
%%%%%%%eq%%%%%%%%
\begin{equation}
\left(R_{x0}^{-1}(\zeta)\right)_{LDA}= 3/2 \ (3/\pi)^{1/3} 
\left( n_0(\zeta) \right)^{1/3} \ \ .
\end{equation}
%%%%%%%endeq%%%%%%% 
%%%%%%%%%%%fig%%%%%%%%%%%
%%%%%%%%%%endfig%%%%%%%%%%%%%
As expected the LDA exchange is much too small for 
$z/l < -1$, but accurate for $z/l \geq 2$. For $\zeta \rightarrow -\infty$, 
$R_{x0}^{-1}(\zeta) \rightarrow \frac{1}{2} \zeta^{-1}$.

\begin{figure}
\begin{center}
\mbox{\epsfig{file=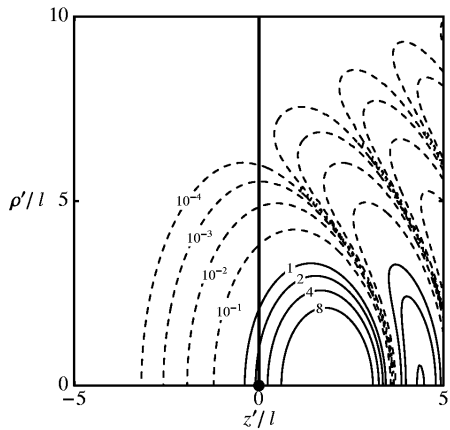,width=2.6in,angle=0}}\\
\end{center}
\caption{The exchange hole density $n_x (\vek{r}')$ of the Airy Gas for $\vek{r}=(0,0,0)$ 
(see the black dot). The vertical line indicates the edge at $z=0$ where $v_{eff}(z)=\mu$. 
The values on the 
contours are given in units of $10^{-3} \, l^{-3}$.}
\label{fig:exchangehole}
\end{figure}

For the AG the exchange hole density has been calculated numerically and
is shown for $\vek{r}=(0,0,0)$ in Fig.~(\ref{fig:exchangehole}). 
%%%%%%%%%%%fig%%%%%%%%%%%
%%%%%%%%%%endfig%%%%%%%%%%%%%
This may be compared with the isotropic exchange hole density of a uniform electron gas 
of density $n$, centered at the point $\vek{r}$ and
given by
%%%%%%%%%%%%eq%%%%%%%%%%%%
\begin{equation}
n_x (\vek{r};\vek{r}') = - 9 \, n \, \, (\sin{\tau}-\tau \cos{\tau})^2/(2 \, {\tau^6})\, ,
\end{equation}
%%%%%%%%%%%%%%endeq%%%%%%%%%%%%%%
where $\tau={(3 \pi^2 n )}^{1/3} | \vek{r}-\vek{r}'|$.

%%%%%%%%%%%%%%%%%%%%%%%%%%%%%%%%%%%%%%%%%%%%%%%%%
Application to the Edge Problem of Thomas Fermi Theory:
For simplicity we first consider electrons with chemical 
potential $\mu$ in a spherical effective potential $v_{eff}(r)$. The edge surface $S$ is a sphere of 
radius $r_e$ given by $v_{eff}(r_e)=\mu$. The ``edge problem'' in TF theory is 
that--incorrectly--$n_{TF}(r) \equiv 0$ for $r>r_e$ (Compare Fig.~\ref{fig:density}). 

Following present ideas we consider points $\vek{r}=(0,0,r)$, where $r \sim r_e$, and their
reference point on $S$, $\vek{r}_e=(0,0,r_e)$. Near $r_e$ we approximate
%%%%%%%%%%%%%%%%%eq%%%%%%%%%%%%
\begin{equation}
v_{eff}(r) = \mu - F \, (r_e-r)\, , \ \ F\equiv | v'_{eff}(r_e) |
\end{equation}
%%%%%%%%%%%%%%%%endeq%%%%%%%%%%%%%%%
which, by spherical symmetry, gives 
%%%%%%%%%%%%%%%%%eq%%%%%%%%%%%%
\begin{equation}
n(r)=l^{-3} n_0((r_e-r)/l) \ \ \ \ \ \ \mbox{(edge region)}
\end{equation}
%%%%%%%%%%%%%%%%endeq%%%%%%%%%%%%%%%
in the edge region which we take to be $r>r_e-\gamma l$ with $l$ defined in 
Eq.~(\ref{eq:ledef}) and $\gamma=2$ (see Fig.~\ref{fig:density}). 
In the interior region, $r<r_e-\gamma l$, we use the standard TF result
%%%%%%%%%%%%%%%%%eq%%%%%%%%%%%%
\begin{equation}
n(r)= \frac{1}{3 \pi^2} \left(\frac{2 m}{\hbar^2} (\mu-v_{eff}(r)) \right)^{3/2}. \  
\mbox{(interior region)}
\end{equation}
%%%%%%%%%%%%%%%%endeq%%%%%%%%%%%%%%%
At $r=r_e-\gamma l$ the two forms approximately agree (See Fig.~\ref{fig:density}).

For an anisotropic $v_{eff}(\vek{r})$, $F(\vek{r}_e)$ ($ \equiv | \nabla v(\vek{r})_{\vek{r}=\vek{r}_e}|$) 
and thereby $l$ will vary over the surface $S$.

Concluding remarks: The concept of ``nearsightness'' allows separate 
consideration of the interior and the edge regions. For the latter 
this paper begins to develop the concept of the edge gas, so far 
without correlation effects. The joining of interior and edge regions, 
using the inverse of the exchange-correlation hole radius, 
$R_{xc}^{-1}(\vek{r})$ (cp. Eq.~\ref{eq:3Dinvraddef}), will be described in a forthcoming work.

Computational support by Dr. F. Pikus and financial support by NSF grant DMR93-0801 and STINT  
are gratefully acknowledged.
%%%%%%%%%%%%%%%%%%%%%%%%%%%%%%%%%%%%%%%%%%%%%%%%%%%%%%%%%%%%

\end{document}